\documentclass[aps,pre,nofootinbib,showpacs,twocolumn,10pt]{revtex4-1}
\usepackage{graphicx,psfrag,color}
\usepackage{amssymb,latexsym,mathrsfs}
\usepackage{amsfonts}
\usepackage{amsmath}
\usepackage{graphicx}

\newcommand{\beq}{\begin{equation}}
\newcommand{\eeq}{\end{equation}}
\newcommand{\beqn}{\begin{eqnarray}}
\newcommand{\eeqn}{\end{eqnarray}}
\newcommand{\bearr}{\begin{array}}
\newcommand{\enarr}{\end{array}}

\newcommand{\ra}{\rangle}
\newcommand{\la}{\langle}

\def\bea{\begin{eqnarray}}
\def\eea{\end{eqnarray}}
\def\ba{\begin{array}}
\def\ea{\end{array}}

\def \theta {\alpha}
\begin{document}
\title{Absorbing State  Phase Transition in presence of Conserved Continuous Local Field}
\author{Mahashweta Basu, Ujjal Gayen and P. K.  Mohanty}
\email[E-mail address: ]{pk.mohanty@saha.ac.in} 
\affiliation {Theoretical Condensed Matter Physics Division, Saha Institute of Nuclear Physics,
1/AF Bidhan Nagar, Kolkata, 700064 India.}  

\begin{abstract}
We study absorbing state phase transition in  one dimension  in presence of a conserved continuous 
local field (CCLF) called energy. A pair of sites on a lattice is said to be active if one or both
sites  posses more energy than a pre-defined threshold. The  active pair of sites  are  allowed
to  redistribute their  energy following a stochastic  rule.  We show that,  the CCLF model  
undergoes a continuous absorbing state transition  when energy per site  is decreased below a critical 
value. The critical exponents are found to be  different from those of DP.  
\end{abstract}  

\maketitle
A state is called absorbing if it is impossible to leave the state. 
Existence of one or more absorbing configurations in a system 
raises a possibility of non-equilibrium phase transition between 
an active and an absorbing state \cite{AAPTBook}. Such absorbing state phase transitions (APT) 
are  encountered 
in a variety of fields, which includes percolation, spreading and 
chemical kinetics \cite{DPBook}.  Numerous physical phenomena, like forest fire \cite{ff},
epidemics \cite{epi}, transport in random media \cite{rm}, synchronization \cite{sync} and  
chromatography can be modeled as APT. The corresponding critical behaviour 
forms a  universality class of APT,  formally known as the 
directed percolation (DP)-class, which  has been  realized 
 \cite{DPexp}convincingly in ($2+1$) dimension.

Long ago, it has been conjectured by  Grassberger and Janssen  \cite{DPconj}  
that  in absence of any special symmetry or conservation law, a continuous phase 
transition into a single absorbing state governed by a  fluctuating scalar order parameter,  
belongs to the DP universality class. Special symmetries, like particle-hole symmetry \cite{u9}, 
conservation of parity  \cite{u10}, and symmetry between different  absorbing 
states  \cite{u11}  lead to different universalities.  Also, 
conserved lattice gas  models  \cite{u14} and conserved threshold transfer 
process (CTTP)\cite{cttp},  where the activity 
field is coupled to the conserved density, show critical 
behaviour different from DP. The DP-conjecture has raised the question, 
`can systems with multiple absorbing states be in DP universality 
class?' In the  pair contact process (PCP), initially introduced by 
Jensen  \cite{pcp}, the  number of absorbing configurations grow to infinity  
with the system size. However, numerical simulations show that 
the critical behaviour  is same as that of DP, which is further supported 
by  phenomenological theories \cite{pcpMunoz}. Other models, like 
the threshold transfer process \cite{m13}, and  dimer reaction ͓
models  \cite{m14}, which have infinitely many absorbing states(IMAS), 
also show static critical exponent same as DP, whereas the 
dynamical exponents (that characterize the spreading of localized 
perturbations) are non-universal  \cite{staticDP}; they  depend on the
nature of the absorbing states and initial conditions\cite{Odor}. Additional conservation laws, 
like parity  \cite{mendes},  can change the  static exponents.

More recent studies of Dickman and coworkers  \cite{fes-soc} have renewed 
the interest  in  APT in systems having IMAS. They were
able to show that the scaling behavior observed in 
sandpile models of  self-organized criticality  \cite{soc}, is entirely governed 
by an underlying absorbing state transition  into IMAS existing in equivalent 
fixed energy sand pile models (FES)\cite{doubt}. Corresponding universal behaviour, which 
are  different from DP, are attributed to the presence of 
coupling of the order parameter to the conserving height field \cite{fes}.
However  certain specific perturbations, like ``stickiness''  \cite{dd-pk}, can 
drive these models to have  critical behaviour same as DP.

In all these models discussed above, the number of absorbing states in a 
finite system  is countable and  grow exponentially to infinity in the thermodynamic limit.
Models with continuous local field variable  may show critical behaviour different from DP as   
(i) there is a possibility of having uncountably infinite number of absorbing states in these systems
(even when the system size is finite) and (ii) as argued by Grassberger \cite{a9}, while
for continuous variables one has ``incomplete death'' discrete systems do not
allow such a process. However, the biological evolution model (BEM)  \cite{lipowsky}, which has a continuous 
dynamical variable,  show DP-critical behaviour. Whereas the {\it coupled map lattice} models,  
where phase transition occurs  to a synchronized (absorbing) phase, follow either  DP or 
Karder-Parisi-Zhang universality class \cite{kpz}, or a first order transition \cite{pkFOT} 
depending on  the non-linearity of 
the {\it map} \cite{pk6}. Exactly which microscopic ingredients
can make an absorbing state transition not belong to the
DP class is an open and challenging problem.

In this article  we  study a  model in one dimensional lattice where  the dynamical variables 
at  the lattice sites, called  {\it energy},   are continuous  and their  sum is conserved. 
They are updated pairwise, when one of the variable crosses a threshold value.  We show that 
the system  undergoes an  absorbing state  phase transition when the  total energy 
crosses a critical value.  Study of several variations of the  models show that the critical 
behaviour of these systems is robust and  forms a new universality class  different from DP.
  
The model  is defined on a one dimensional lattice  with periodic boundary condition with 
sites  labeled by $i=1,2\dots L.$     
The dynamical variable  called energy $E_i$,  at each site  $i$, is continuous 
and satisfies $E_i= E_{i+L}$  (periodic boundary condition).   A pair of neighbouring 
sites are called {\it active} when  energy of one or both the sites cross a threshold 
value $w$. Otherwise,  $i.e.$  when both sites have energy less than $w$, they 
are \textit{inactive}.  Only the active pairs are allowed to exchange energy through 
an energy conserving dynamics,
\bea
E_i &\to& \lambda E_i  + \epsilon (1-\lambda) (E_i+E_j)\cr
E_j &\to& \lambda E_j  + (1-\epsilon) (1-\lambda) (E_i+E_j),
\label{eq:cc}
\eea
where $\lambda$ is a parameter of the model and $\epsilon$  is a random number 
distributed uniformly in $(0,1)$. Clearly the total energy  $E=\sum_i E_i$ is 
conserved by this dynamics.  

A special case of the model,  $w=0$ corresponds 
to the kinetic model of markets introduced by Chakraborti and Chakrabarti (CC) \cite{cc} in the 
literature of econophysics, where $E_i$ is considered as wealth of an agent $i$  who can interact 
with any other agent $j$ (not necessarily  its neighbour). Naturally, in this mean-field model 
the parameter $\lambda$  denotes the savings propensity. Variations of these models with 
variable savings propensity \cite{CCM}, provide the first  explanation \cite{pkCCM}  
`why tails of the wealth distribution follow a power law called Pareto law \cite{Pareto}'.
Again, when $\lambda=0$, the distribution of  energy  (or wealth)  follows a 
Gibb’s distribution \cite{yv}  which has been observed in distribution of income-tax return of
individuals in several countries \cite{tax}.

In  fact  $\lambda=0$ case has been studied  earlier \cite{Kipnis} on a lattice, in context of heat conduction, 
as a model of collisional dynamics of particles.  First we consider this case $\lambda=0$  with a finite 
threshold $w$ (taken as $w=1$, without loss of generality). Note, that  a different  threshold $w\ne1$ 
would  shift the transition point linearly.

To study the properties of this model, particularly   possibility of phase transition and  critical behaviour,
we used Monte Carlo simulations.  First let us define local activity field $s_i=0$   when the neighbouring 
sites  $i$ and $i+1$ are inactive, $i.e.$ when  both the  sites have less than unit amount of energy. Or otherwise 
$s_i=1$ (the {\it bond} joining $i$ and $i+1$ is active).  Clearly, the activity (or the 
energy exchange) does not die 
out when energy density $e=\frac 1 L \sum_i E_i$ is larger than $1$, as in this case the system   has at least one active  bond;    
starting from any arbitrary configuration, the density of active  bonds $\rho(t) = \la s_i\ra$, reaches 
a stationary value $\rho_s$ as $t\to \infty$. Whereas for small energy density  $ e\ll1$,  the activity 
is  expected  to die out,  because  it is highly improbable to have macroscopic number of sites with energy larger 
than $w$.  This indicates that there  may be an absorbing state phase transition at $e_c<1$. 

\begin{figure}[h]
 \centering
 \includegraphics[width=7cm]{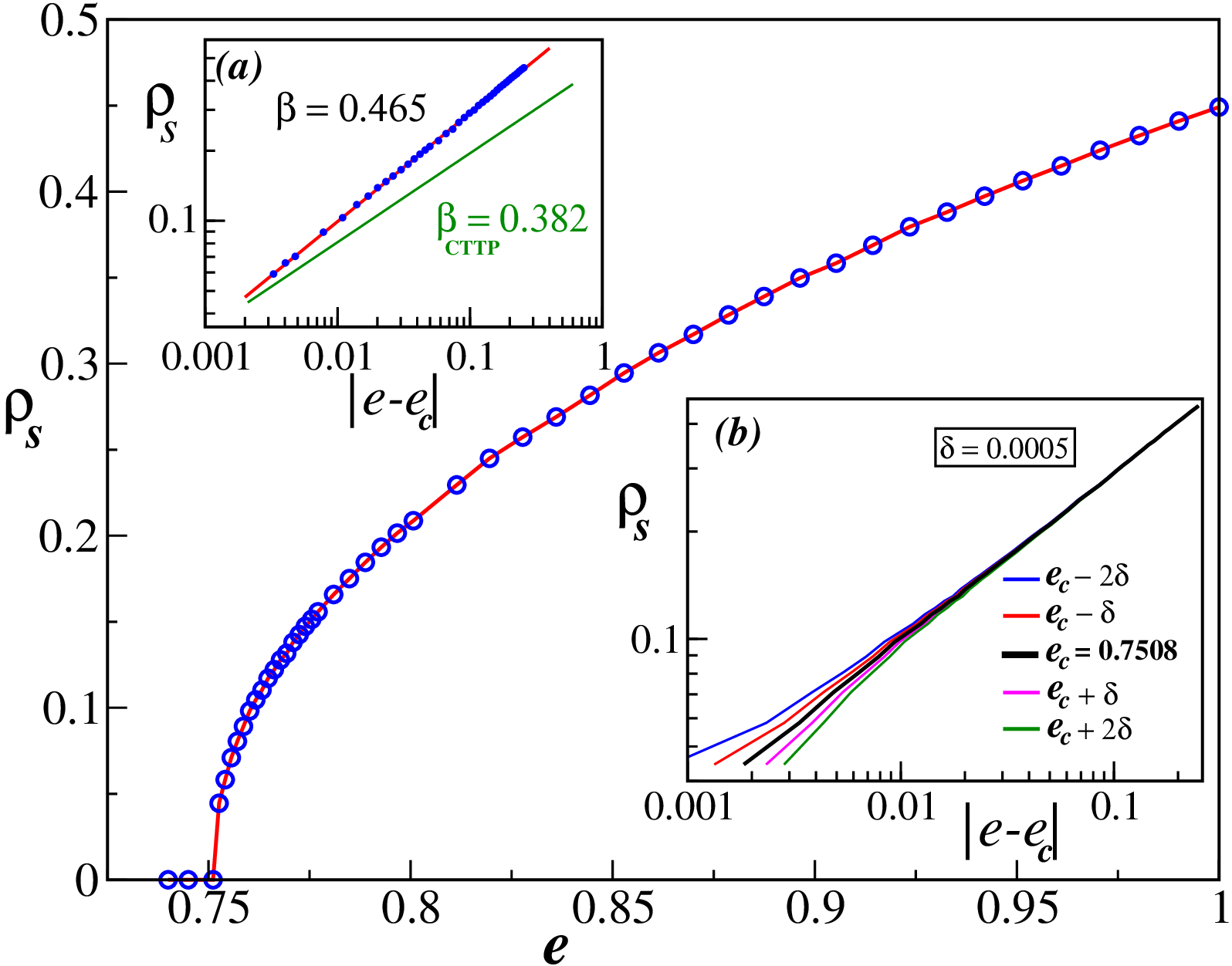}
 \caption{(Color online) The order parameter $\rho_s$ as a function of $e$.  
In log scale,  $\rho_s$ is found to be linear in  $(e-e_c)$ by choosing  $e_c=0.750(8).$
Any  small change  (in steps of $\delta = 0.0005$) gives  deviation from linearity (see inset (b)).
Inset (a) shows   that the orderparameter vanishes as $\rho_s\sim (e-e_c)^\beta$,  
with $\beta = 0.465$, which is visibly different  from   $\beta_{CTTP} = 0.382.$ 
Here, the system size is $L=10^4.$ 
 }
 \label{fig:beta}
\end{figure}
In Fig. \ref{fig:beta}, we have plotted  the steady state  density of active sites $\rho_s$ as a function 
of  energy density $e.$  It is evident that $\rho_s$  takes a nonzero value  as $e$ crosses $e_c$. 
$\rho_s$ is plotted  in log scale as a function of $(e-e_c)$ show a straight line  by choosing 
the critical value $e_c= 0.750(8)$ (see Fig.  \ref{fig:beta}(b)).   The orderparameter exponent 
$\beta=0.46(5)$,  defined from the relation  $\rho_s\sim (e-e_c)^\beta$,  is clearly  different 
(Fig.\ref{fig:beta}(a))  from that of  $\beta_{CTTP} =0.382$. 

\begin{figure}[h]
 \centering
\includegraphics[width=7cm]{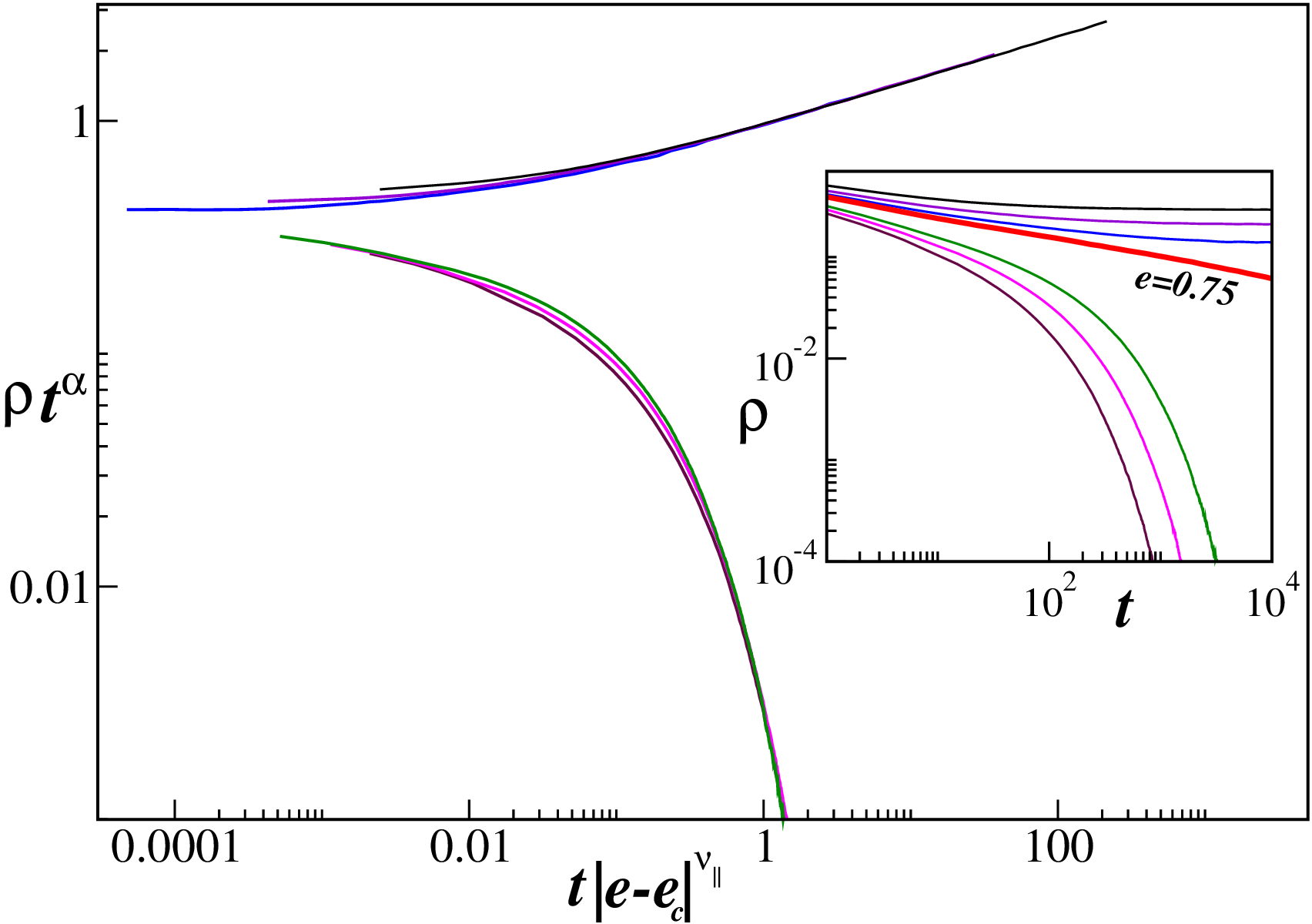}
 \caption{(Color online)   Inset : Decay of $\rho(t)$ from a fully active state to the  
steady state value $\rho_s$ is shown in the 
inset for $L=1000$ and $e=0.65, 0.67, 0.69, 0.75, 0.77,0.8,$ and $0.85$ .  
Main figure : Data collapse for off-critical simulations according to 
the scaling form Eq. \eqref{eq:rhot_eps} is obtained for $\theta=0.19(5)$  and 
$\nu_\parallel=2.64.$ }
 \label{fig:theta}
\end{figure}

One can obtain  a few other exponents from the decay of  $\rho(t)$ from a
fully active configuration\footnote{This can be generated  by taking an initial
condition such that  all sites of  one sub-lattice 
have energy larger than unity.} with $\rho(0)=1.$ After an initial decay  $\rho(t)\sim t^{-\theta}$ it
approaches the steady state value $\rho_s$ in  the $t\to\infty$ limit. So $\rho$  must scale as
\bea 
\rho(t, e) = t^{-\theta} {\cal F}\left(t^{1/\nu_\parallel} (e-e_c) \right).
\label{eq:rhot_eps}
\eea 
Thus, one expects that $\rho(t)$ for different values of $e$  (shown in  the Fig. \ref{fig:theta}) collapse into a 
single scaling function ${\cal F}$, when $t^{\theta}\rho(t)$ is plotted   against $t |e-e_c|^{\nu_\parallel}$ in 
log scale. The main figure here shows the data collapse when we use $\theta=0.19(5)$, $ \nu_\parallel=2.64.$  
Since at the critical point $\rho(t, e_c)=  t^{-\theta} {\cal F}(0)$, one can obtain both $e_c$ and $\theta$ directly.  
Resulting $e_c$ and $\theta$ are consistent with those obtained from the  data collapse.
Again, in $t\to \infty$ limit,  $\rho$ vanishes  as $|e-e_c|^\beta$. This can happen only when  
${\cal F}(x)\sim x^{\beta/\nu_\parallel}$; thus  $$\theta = \beta/\nu_\parallel.$$
Since all three exponents $\beta$, $\theta$ and $\nu_\parallel$ are calculated independently, one can check if 
the above scaling relation holds. In fact, it holds to a great accuracy for the values of 
$\beta$, $\theta$ and $\nu_\parallel$ calculated here.  

\begin{figure}[h]
 \centering
 \includegraphics[width=7cm]{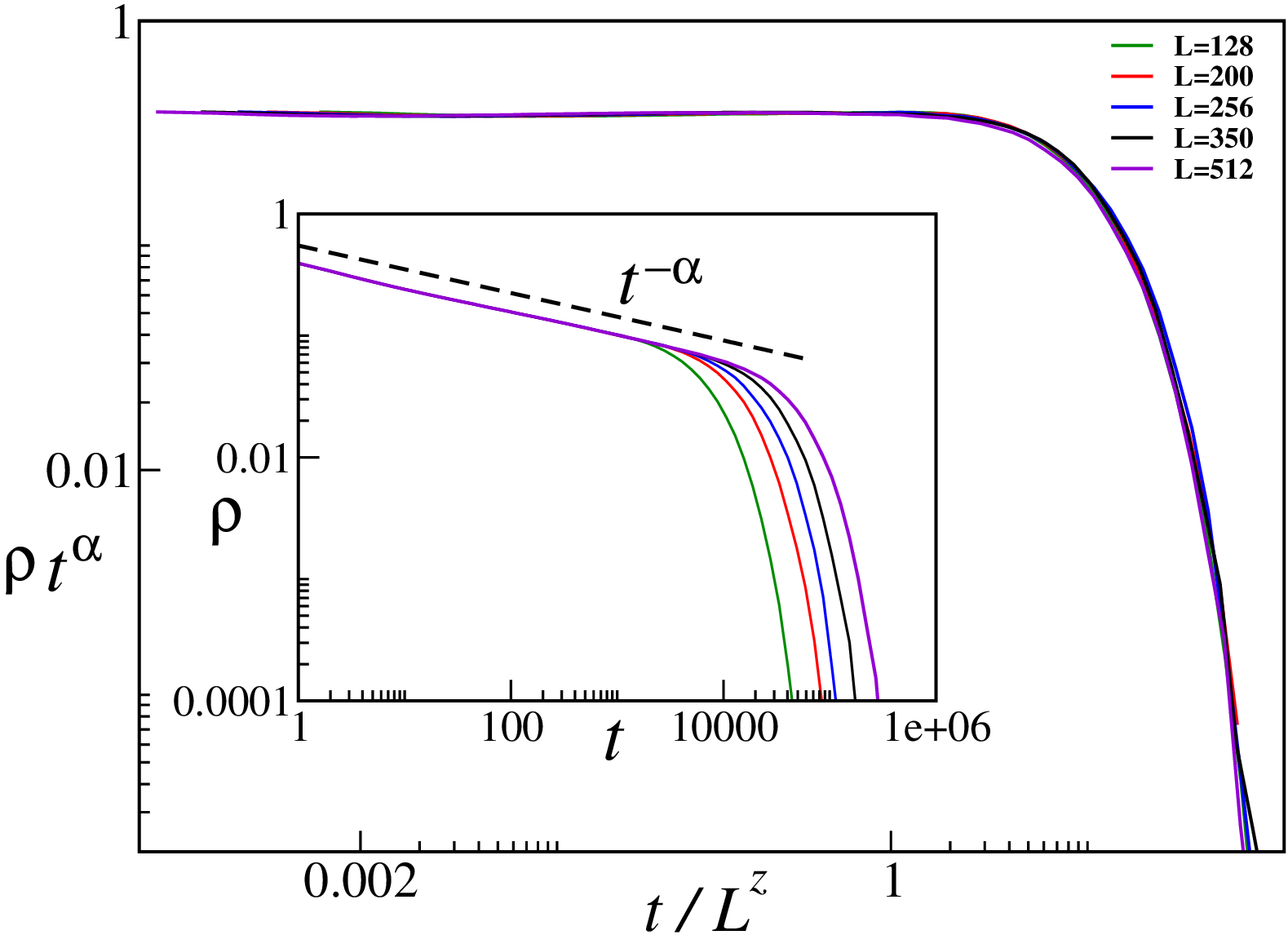} 
 \caption{(Color online) The  decay of $\rho(t)$  at $e_c=0.750$ from a fully active state, averaged over 
$10^4$ ensembles,  is shown in the inset for different $L=128,200,256,350,$ and $512$.    A line with 
slope $\theta=0.195$ is  drawn to emphasize that 
$\rho(t) \sim  t^{-\theta}.$ The main figure shows the  data collapse for finite size scaling 
Eq. \eqref{eq:rhot_L}  using $z=1.381.$ }
 \label{fig:z}
\end{figure}

Now we turn our attention to the finite size scaling of $\rho(t)$  at the critical point. Again,
from a fully active state $\rho(t,L)$ decays as $t^{-\theta}$, indicating a scaling form
\bea 
\rho(t, L) = t^{-\theta} {\cal G}(t/L^z),
\label{eq:rhot_L}
\eea 
where $z$ is the dynamic exponent. Thus,  one expects that $\rho(t)$ for different values of $L$ collapse
to a single function when plotted against $t/L^z$.  This is described in Fig. \ref{fig:z}.
The inset there shows variation of $\rho(t)$ for different $L=128,200,256,350,512$, which were made to collapse 
to a single function using $\theta=0.19(5)$ and $z=1.38(1)$. Now, assuming the scaling relation 
$z= \nu_\parallel/\nu_\perp,$ one can obtain $\nu_\perp = \beta/\theta z.$

The exponents we obtained  using Monte-Carlo simulations are  listed in Table \ref{table}, along with the 
exponents of DP, BEM and CTTP.  Clearly,  the exponents of CCLF model are quite different
from the DP. Even the  static exponents, which are
believed \cite{staticDP} to be same as that of DP for models with multiple absorbing sates,  are now different.
Though some of the  exponents are  close to  the same obtained for CTTP, 
particularly the order parameter exponent $\beta$, and the spreading exponent $\alpha$  are very different.  
Since the value of $\beta$ crucially relies on the estimation of  the critical point, we  provide a careful 
study of $e_c$ based on Ref. \cite{note1} in Appendix-I. From this analysis it is concluded,  
beyond reasonable doubts, that $\beta=0.46(5)$  and   thus the phase transition in CCLF model 
belong to a new universality class.
 
One can possibly reason it to  the existence of a continuous conserved field. Further study
in this direction is required to identify, what made this absorbing state phase transition
different from the usual ones.     
\begin{table}[h]
\caption{Critical exponents of DP, BEM  and CCLF  model.}
\begin{tabular}{ccccc} 
\hline\hline
 &$\beta$&$\nu_\parallel$&$\theta$& $z$\\ \hline
DP \cite{DPBook} & 0.276 & 1.733&0.159&  1.581\\
BEM \cite{lipowsky} & 0.276 & 1.067&.259&1.364\\
CCLF & 0.46(5) &2.6(4) &  0.19(5)&1.38(1) \\
CTTP \cite{cttp} &0.382 &2.45&0.141 & 1.393
\\ \hline\hline
\end{tabular}
\label{table}
\end{table}

In the following we discuss some possible directions of studies which may  explain 
the new universality. The model studied here, is quite similar to the  sandpile models \cite{soc} 
of self organized criticality, except that the local field variables in CCLF model are continuous.
It has been pointed out
in Ref.  \cite{dd-pk},  that although the critical behaviour of  sandpile models  crucially depends 
on the details of the dynamics and the spatial dimension,  they are unstable to certain  specific 
perturbation (called  {\it stickiness}) and flow  generically to  the DP universality class. It would be 
interesting  to ask, if the model studied here  is unstable to perturbations.   Some generic 
variations, which may change the critical behaviour of CCLF, are discussed below.

The bulk dynamics of CC model ($\lambda \ne 0$) does not  satisfy 
detailed balance, as  for any arbitrary $(E_i,E_j) \to (E_i^\prime, E_j^\prime)$ it is 
not possible to have  $(E_i^\prime, E_j^\prime) \to (E_i,E_j).$  Thus, $\lambda \ne 0$ is 
a singular perturbation and it may change the universality class. The quenched disorder, 
introduced   by taking a distributed savings propensity $\{ \lambda_i\}$ \cite{CCM} may 
also change the critical behaviour.

In conclusion, we have studied absorbing state phase transition in a model with conserved continuous local field (CCLF). 
In one dimension, the  model is defined on a lattice with  sites $i$ having continuous variable $E_i$ called energy. 
Two neighbouring sites $i$ and $i+1$ can exchange energy when  one or both sites have energy  larger than a 
predefined threshold $w.$  The exchange dynamics is  similar to the wealth exchange model  \cite{cc} studied earlier, 
by considering $E_i$ as wealth of the agent $i$.  Clearly, the system is active ({\it i. e.}  surely some of the neighbouring 
sites  keep  exchanging their wealth)  when the energy  density $e$ is larger than $w$ (set to be unity, without loss 
of generality). Whereas  for $e\ll 1$  it is improbable to get  macroscopic number of sites which have wealth larger 
than $w$ indicating that the system falls into  one of the  uncountably infinite number of absorbing states. 
We show that, CCLF model undergoes a continuous phase transition  at $e_c= 0.750(8)$ with critical exponents  
$\beta=0.46(5)$,  $\nu_\parallel=2.6(4)$,  $\theta=0.19(5)$, and $z=1.38(1)$, which are 
very different from those of DP.  However earlier studies \cite{staticDP},  both numerical and  analytical,
have shown that the static exponents of  absorbing state   transition to infinitely many absorbing  
states are  same as those of DP, whereas dynamic exponents differ.  It is surprising, that  in CCLF model,
even the static exponents differ from DP and form a new universality class. One may argue that the existence of  the conserved field  is responsible for this behaviour.
 But, this can not be  the sole argument as recent studies \cite{dd-pk} show  that  it is possible 
to get DP behaviour in presence of conserved field(s).
\smallskip

\centerline{\appendix{\bf Appendix-I}} \smallskip

The numerical estimation of the critical exponents  of  CCLF model are listed in Table-I. 
The critical exponents $\nu_\parallel$ and $z$ are quite close to that of CTTP, whereas $\beta$ and 
$\theta$ are substantially different. The main claim of this article, that CCLF model 
form a new universality class, is based  on this difference. 
Since the value of $\beta$  strongly depends on the the critical point , 
here we  estimate $e_c$  through a 
careful  and systematic numerical analysis based on Ref. \cite{note1}. 
\begin{figure}[h]
 \centering
 \includegraphics[width=8.5cm]{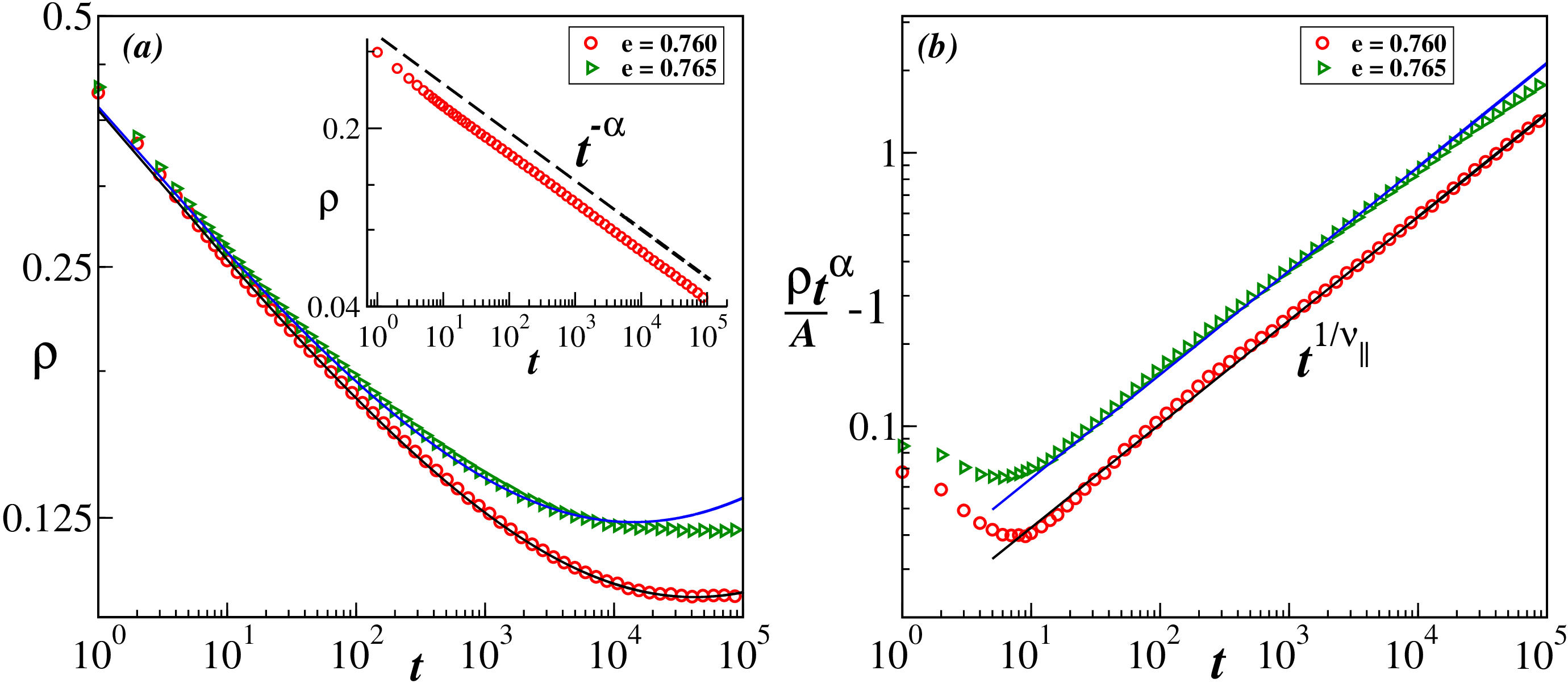}
 \caption{(Color online) (a) $\rho$ versus  $t$  for $e=0.76$ and $0.765$ are fitted with the scaling form  \eqref{eq:form}. 
(b) $A^{-1}\rho t^\theta -1$ as a function $t$ in log scale  is linear  for both $e=0.76$ and $0.765$ when $A=2.64$; the solid lines indicates a power-law $t^{1/\nu_\parallel}$ with $\nu_\parallel=2.64$  and  coefficients $0.0178 $ and $0.0269$ respectively. From these coefficients  we find that $B=1.82$ and $e_c= 0.75022.$  The inset of (a) shows that 
$\rho(t,e_c)$ obtained from $\rho(t,e=0.76)$  using Eq. \eqref{eq:form1} varies as $t^{-\theta}$ with $\theta=0.19$,  when  
$e_c$ is taken as $0.7503.$  Here  $L=1024$.}
 \label{fig:ec}
\end{figure}

Near the critical point, the order parameter $\rho(t,e)$  follow Eq. \eqref{eq:rhot_eps}; thus  
\bea
\rho(t,e)&=& A t^{-\theta}  (1+ B(e-e_c) ~t^{1/\nu_{\parallel}} )\cr
         &\simeq & \rho(t,e_c) ( 1  + B(e-e_c) ~t^{1/\nu_{\parallel}} ) \label{eq:form}
\eea
where  we  have used the  Taylor's  expansion of scaling function ${\cal F}(x)$  at $x=0$ upto the  first term, 
$i. e.~ {\cal F}(x) = A(1+ B x).$   Clearly, this functional form is valid for $t \ll  |e- e_c| ^ {-\nu_\parallel}.$

Using this functional form, which  is valid for $t \ll  |e- e_c| ^ {-\nu_\parallel},$ one can obtain
$\nu_\parallel$   by suitably choosing $A$ such that $A^{-1}\rho(t,e)t^{\theta}-1$,  as a function of $t$,   
is linear in log scale; the slope and the $y$-intercept determine the  $\nu_\parallel$ and  $B(e-e_c)$ respectively. 
For a system of size $L= 1024$,  $\rho$ is shown in  Fig. \ref{fig:ec} (a) for two different values 
of $e=0.76$ and $0.765$.  Figure \ref{fig:ec}(b)  shows that $A^{-1}\rho(t,e)t^{\theta}-1$ is linear in log
scale for both values of $e$, with  $\theta=0.19$ and $A^{-1} = 2.64$.  The slope of both the straight-lines 
turns out to be $1/\nu_\parallel=0.379$, which is consistent  with $\nu_\parallel=2.6(4)$ obtained earlier from  the data-collapse. From their $y$-intercepts we get, $B(e-e_c)= 0.0178,0.0269$ respectively for 
$e=0.76$ and $0.765,$   which  provide $e_c= 0.75022$ and $B=1.82.$  Using these parameters, we have calculated 
$\rho(t,e)$   as a function of $t$  for $e=0.76,0.765$ (shown as solid lines in  Fig.  \ref{fig:ec}(a)).  The
discrepancy for large $t$, in case of $e=0.765$ is due to the fact that we have  approximated  ${\cal F}(x)$ in Eq. \eqref{eq:form} up to the linear order.
  
A better estimation of $e_c$ can be done now by using   Eq. \eqref{eq:form}, 
\bea
\rho(t,e_c)=\frac{ \rho(t,e)}{  1  + B(e-e_c) ~t^{1/\nu_{\parallel}} }. \label{eq:form1}   \eea
Since we know the value of $B$ and   $\nu_\parallel$, $e_c$ can be used as a fitting parameter  in the above 
equation such that  $\rho(t,e_c)$ obtained from  the numerical values of $\rho(t,e)$ using 
Eq. \eqref{eq:form1}
shows a power-law.  The inset of Fig.  \ref{fig:ec} (a) shows that  $\rho(t,e_c)$, obtained from  
$\rho(t,e=0.76)$  by choosing $e_c=0.7503$, decays algebraically  for  four decades  with exponent $\theta=0.19.$
Thus the final estimate of  the critical point for $L=1024$ is $e_c= 0.7503.$


\begin{figure}[h]
 \centering
 \includegraphics[width=8.5cm]{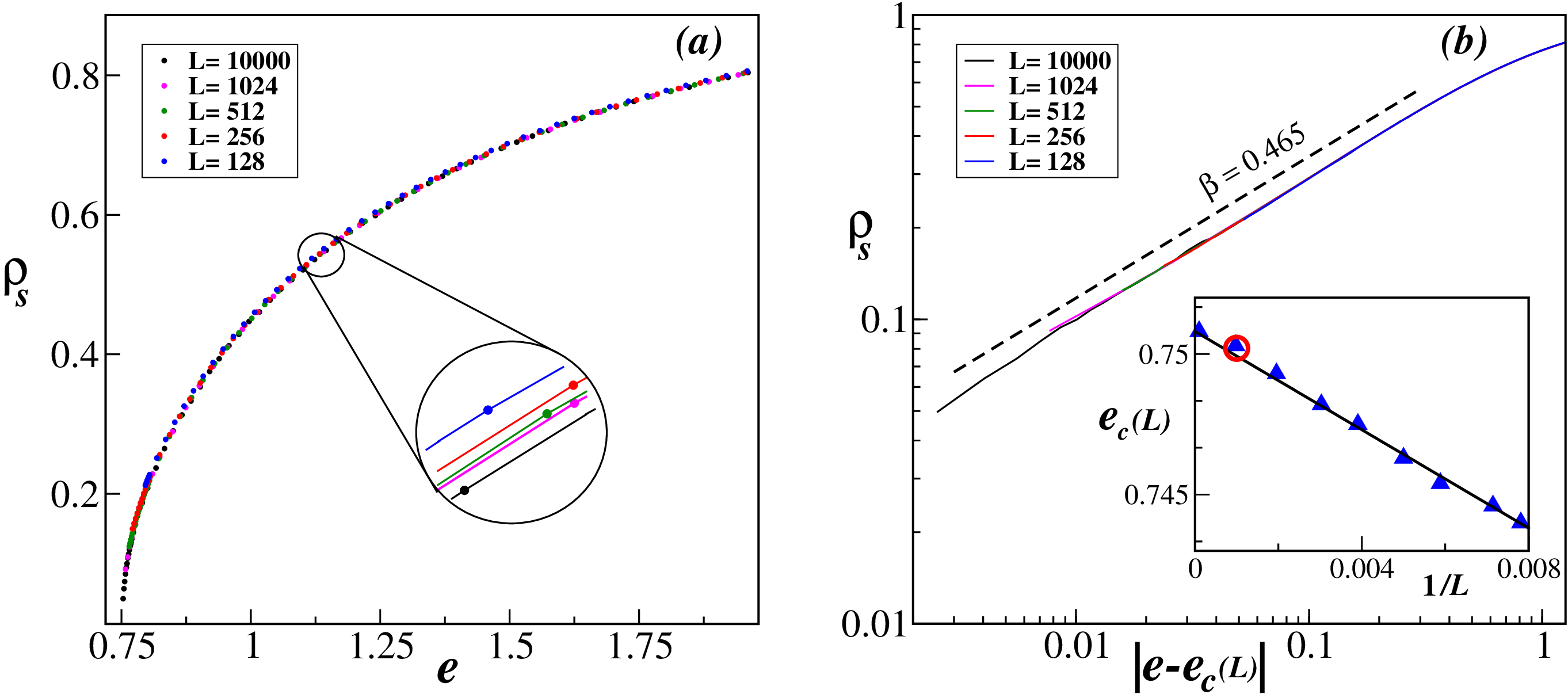}
 \caption{(Color online) (a) $\rho_s$ as  a function of $e$ for different $L=128,256,512,1024$ and $10^4$.  Difference in $\rho_s$ 
is visible only after sufficient magnification. (b)  $\rho_s$  for different $L$  are merged onto  
the same for $L=10^4$  by using Eq. \eqref{eq:ecL}. The inset here show variation of $e_c(L)$ with $1/L$, which is linear. The critical value obtained for $L=1024$ following  Ref. \cite{note1} is shown as a circle. The asymptotic value of the critical point, extrapolated for 
$L\to \infty$ is  $e_c=0.7508.$ }
 \label{fig:ecL}
\end{figure}

In general, the order parameter and thus the critical point  depends  strongly on system size  
$L$.  In Fig. \ref{fig:ecL} (a), we  have  plotted $\rho_s$ as a  function of $e$, for 
$L=128,256,512,1024$ and $10^4$.   These curves  do not show any significant dependence on $L$. 
The small difference can be adjusted by shifting the $x$-axis by a small amount $\delta_L$, which 
results in 
\bea\rho_s^L(e) = \rho_s^{\infty}(e+ \delta_L)  \sim (e-e_c(L) )^\beta,  \label{eq:ecL}\eea
where $e_c(L) =  e_c(\infty)  - \delta_L.$  We calculate $e_c(L)$ as the value, where the log scale plot 
of $\rho_s$ versus  $e-e_c(L)$  is linear.  In fact, with these choice of $e_c(L)$, all the  different curves 
of  $\rho_s^L$ could be merged on to  the curve for $L=10^4$ (see  Fig.  \ref{fig:ecL} (b)).  The inset of 
Fig.  \ref{fig:ecL} (b) shows  the small variation of $e_c(L)$ with $1/L$, which asymptotically approach 
to $e_c=0.7508$ as $L\to \infty.$  
Again, the value of $e_c$ obtained in this Appendix for $L=1024$,  using  a method described in Ref. \cite{note1}, consistently fall on this curve (denoted by a circle in the inset of 
Fig. \ref{fig:ecL} (b)) .  

 These analysis suggest that  the estimates of $e_c$  and the critical exponents $\beta$ and $\alpha$ 
{\it do not} change appreciably with system size. The current values of exponents obtained for 
system size $L=10^4$   are sufficient to conclude that the  critical behaviour in CCLF model 
belongs a universality class different from  CTTP.


\end{document}